\documentclass[a4paper, amsfonts, amssymb, amsmath, reprint, showkeys, twoside, superscriptaddress,amsart]{revtex4-1}
\usepackage[english]{babel}
\usepackage{mathrsfs}
\usepackage{amsmath,amssymb,graphicx}
\usepackage[utf8]{inputenc}
\usepackage[colorinlistoftodos, color=green!40, prependcaption]{todonotes}
\usepackage{amsthm}
\usepackage{mathtools}
\usepackage{physics}
\usepackage{xcolor}
\usepackage{graphicx}
\usepackage[left=23mm,right=13mm,top=35mm,columnsep=15pt]{geometry} 
\usepackage{adjustbox}
\usepackage{placeins}
\usepackage[T1]{fontenc}
\usepackage{lipsum}
\usepackage[pdftex, pdftitle={Article}, pdfauthor={Author}]{hyperref}
\begin{document}

\title{Weak thermal state quadrature-noise shadow imaging}
\author{Pratik J. Barge\footnote{author footnote}}
\email[]{barge.pratik@gmail.com}
\affiliation{Hearne Institute for Theoretical Physics, and Department of Physics and Astronomy,
Louisiana State University, Baton Rouge, Louisiana 70803, USA}
\author{Ziqi Niu}
\affiliation{Department of Physics, William \&
Mary, Williamsburg,
Virginia 23187, USA}
\author{Savannah Cuozzo}
\affiliation{Department of Physics, William \&
Mary, Williamsburg,
Virginia 23187, USA}
\author{Eugeniy E Mikhailov}
\affiliation{Department of Physics, William \&
Mary, Williamsburg,
Virginia 23187, USA}
\author{Irina Novikova}
\affiliation{Department of Physics, William \&
Mary, Williamsburg,
Virginia 23187, USA}
\author{Hwang Lee}
\affiliation{Hearne Institute for Theoretical Physics, and Department of Physics and Astronomy,
Louisiana State University, Baton Rouge, Louisiana 70803, USA}
\author{Lior Cohen\footnote{author footnote}}
% \thanks{Currently at University of Colorado Boulder}
\email[]{lior.cohen3@mail.huji.ac.il}
\affiliation{Hearne Institute for Theoretical Physics, and Department of Physics and Astronomy,
Louisiana State University, Baton Rouge, Louisiana 70803, USA}
\affiliation{Department of Physics, University of Colorado, Boulder, CO 80309, USA}
% \email[Correspondence email address: ]{lior.cohen3@mail.huji.ac.il}

\date{\today}

\begin{abstract}
In this work, we theoretically and experimentally demonstrate the possibility to create an image of an opaque object using a few-photon thermal optical field.  We utilize the Quadrature-Noise Shadow Imaging (QSI) technique that detects the changes in the quadrature-noise statistics of the probe beam after its interaction with an object. We show that such thermal QSI scheme has an advantage over the classical differential imaging when the effect of dark counts is considered. At the same time, the easy availability of thermal sources for any wavelength makes the method practical for broad range of applications, not accessible with, e.g. quantum squeezed light. As a proof of principle, we implement this scheme by two different light sources: a pseudo-thermal beam generated by rotating ground glass (RGG) method and a thermal beam generated by Four-Wave Mixing (FWM) method. The RGG method shows simplicity and robustness of QSI scheme while the FWM method validates theoretical signal-to-noise ratio predictions. Finally, we demonstrate low-light imaging abilities with QSI by imaging a biological specimen on a CCD camera, detecting as low as 0.03 photons on average per pixel per 1.7~$\mu$s exposure.
\end{abstract}


\maketitle

\section{Introduction}
We routinely reconstruct shapes of various objects by shining light on them and then analyzing the transmitted intensity. The sensitivity of such measurement is determined by intrinsic fluctuations of the photon number in the light field (the shot noise limit in case of a coherent probe beam with Poissonian statistics~\cite{meda2017photon,berchera2019quantum}), as well as by technical issues, such as the dark noise of a CCD camera~\cite{konnik2014high}, stray light leakage, etc. For a bright probe with $10^{3} - 10^{5}$ photons per pixel these noise sources are often negligible~\cite{morris2015imaging}. However, many applications such as imaging fragile samples, biological specimens, photosensitive chemicals, or performing covert military operations require operation at much lower illumination levels. In this case useful signals may be overwhelmed by the noise, putting stringent requirements on acquisition time, camera dark noise level, etc. Multiple approaches, such as absorption measurements with non-classical optical fields~\cite{moreau2017demonstrating,whittaker2017absorption,losero2018unbiased} or quantum correlation analysis~\cite{brida2010experimental,ferri2010differential,samantaray2017realization,taylor2016quantum,morris2015imaging} have successfully proven the enhancement in the imaging sensitivity over classical methods using quantum resources such as spatial correlations in twin-beam light produced by spontaneous parametric down conversion (SPDC) process or four-wave mixing in atomic vapor \cite{clark2012imaging,marino2012extracting}. In this work we demonstrate how an optical field with thermal statistics can be used to image an opaque object with only a few photons per image, while dramatically reducing sensitivity to technical noises.

\begin{figure*}[ht]
    \centering \includegraphics[width=\textwidth]{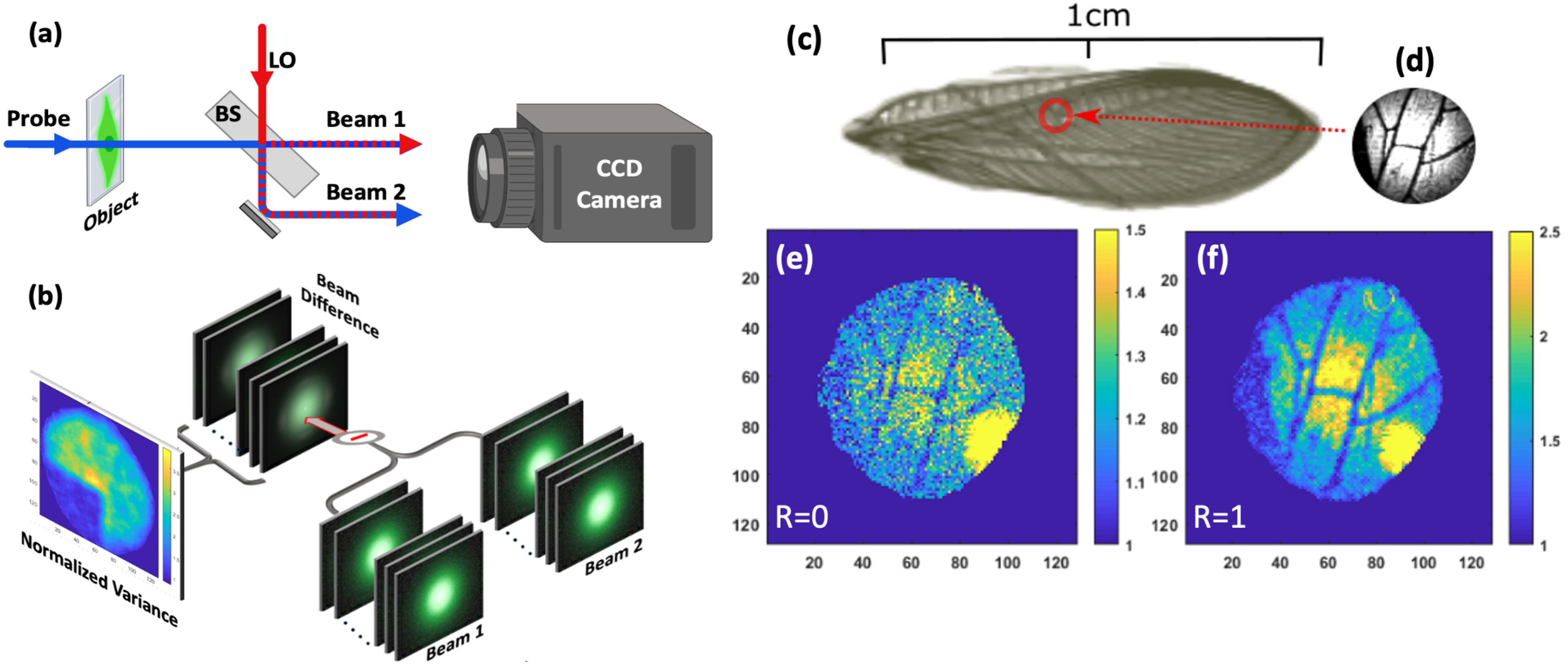}
    \caption {(a) Schematic of the QSI method. Probe field illuminates the object to be imaged and the transmitted part of it is mixed with the local oscillator at a 50:50 beam splitter and captured on a CCD camera. (b) Transmission map of the object emerges as the mean temporal variance of the difference of two output beams is calculated. (c) Image of an insect wing with dimension indicated for reference. (d) Inset shows enlarged image of the region of interest, obtained with bright beam. (e) Normalized variance map of the wing obtained with $\approx$301,600 photons and without any processing. The resolution is around 10 $\mu m$ and is limited by the optical setup and can be further improved. (f) Normalized variance map after binning (see Appendix \hyperref[appendix:binning]{B}) all the pixel values within one pixel radius i.e. $R=1$. The contrast of image is increased at the expense of spatial resolution. We note that a thermal state with $0.1$ photons/pixel/exposure on average is used for generating image (e). See Fig. B1 in Appendix \hyperref[appendix:binning]{B} for images generated with $0.03$ photons/pixel/exposure on average.}
    \label{Fig:schematic}
\end{figure*}

Instead of fighting the photon quantum noise, QSI uses it as a useful resource. As the imaged object blocks portions of the probe optical field, it replaces the original quantum state with coherent vacuum. Thus, the object shape can be reconstructed by detecting local changes in the quadrature noise of the transmitted probe beam. While the probe itself may contain only a few photons, at the detection stage it interferes with a strong classical local oscillator to amplify its quadrature noise and effectively eliminate detrimental technical effects, such as the camera dark noise or background leaks. It is essential for the QSI probe field to have non-Poissoinian statistics, so its quadrature noise is distinguishable from the coherent vacuum. For example, a prior experiment demonstrate image reconstruction using a squeezed vacuum probe field with overall less than 800 photons, with average less than one photon per frame~\cite{cuozzo2022low}. However, thermal probe potentially allows for better resolution compared to the squeezed vacuum probe, which in the latter case is limited by the mode size of the squeezed vacuum state. On the contrary, a portion of the thermal field still displays the thermal statistics with lower photon number, so the spatial resolution can be optimized depending on the required SNR. Moreover, the applicability of a squeezed vacuum probe is less practical due to limited availability and technical challenges in its generation. 

Thermal light sources, on the other hand, are widely available for any spectral region, making them ideal choice for a ``noisy'' QSI probe. In this paper we evaluate, both theoretically and experimentally, the prospects of using thermal optical states with low average photon numbers for QSI imaging. First, we theoretically demonstrate that in the presence of the camera dark noise thermal QSI can outperform classical absorption imaging. Then we confirm this experimentally by producing high-quality images of a complex objects using two low-intensity light sources. The first one is a pseudo-thermal light generated by passing a weak laser beam through a rotated ground glass~\cite{li2020photon} to demonstrate the versatility of the proposed approach for any desired wavelength. The second set of measurements uses a true thermal light generated in the FWM process~\cite{bachor2019guide,mccormick2008strong,boyer2008entangled} for precise comparison with the theoretical predictions (since its properties were well characterised in previous research).

\section{Theoretical Model}
\label{sec:II}
In this section we describe the detailed protocol of the QSI scheme in terms of density matrix. This is a general formalism that allows to analyze the performance of any kind of probe state, both quantum and classical. A low intensity probe field illuminates the object to be imaged. The object transmits some part of it and scatters the rest. Noise statistics of the transmitted field is analysed with a homodyne-like scheme. This is achieved by letting the transmitted field interfere with a strong local oscillator (LO) on a balanced beam splitter and consequent detection by a CCD camera \cite{matekole2020quantum}. Intensity maps of two output beams, recorded at multiple time instants by the camera, are subtracted to get a series of "beam difference" maps as shown in Fig.~\ref{Fig:schematic}(b). A temporal variance map of this time series data, normalized by the total intensity, is the signal for this imaging scheme. Refer to Appendix \hyperref[appendix:FWM]{C} for the details of experimental procedure). Since this scheme employs quadrature variance of the probe state, essentially any state with quadrature variance different from coherent state should work as a probe. Thermal state with its super-Poissonian photon statistics and simple, cost-effective producibility is one such notable candidate. Moreover, availability of thermal probes in wide range of wavelengths makes them promising, particularly for imaging applications.

\subsection{Calculation of Normalized Variance}
\begin{figure}[ht]
    \centering
    \includegraphics[width=\columnwidth]{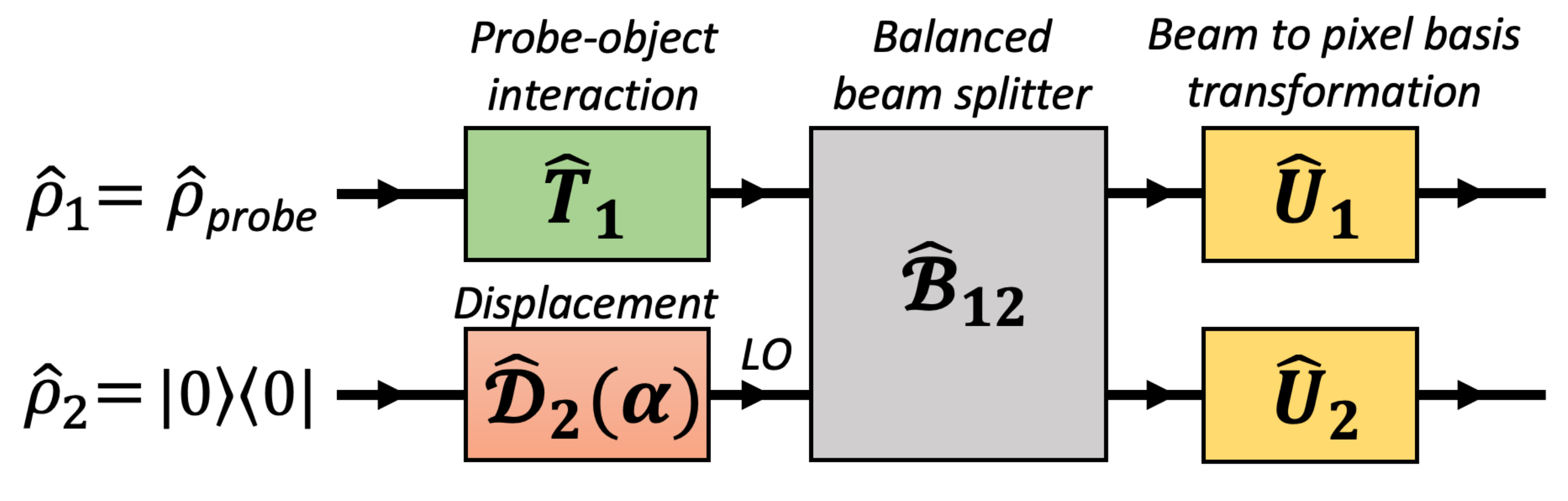}
    \renewcommand{\figurename}{Figure}
    \caption {Block diagram of operator actions in QSI scheme: $\hat{T}_1$ acting on probe state $\hat{\rho}$ in mode 1 represents the object-probe interaction. In mode 2, operator $\hat{D}_2(\alpha)$ displaces vacuum state. Resulting states in both modes are allowed to interfere by the beam splitter operator $\hat{B}_{12}$. Two output modes of beam splitter are transformed from the beam basis to the pixel basis by the mode transformation operators, $\hat{U}_1$ and $\hat{U}_2$.}
    \label{operators}
\end{figure}

Fig.~\ref{operators} represents the theoretical model we use to model the imaging process. A single-mode thermal probe $(\hat{\rho_{1}})$ with $\expval{\hat{n}_{th}}$ of average photons, interacts with an object to be imaged $(\hat{T}_1)$ in mode 1 and mixed with a strong, mode-matched LO $(\hat{D}_2(\alpha)|0\rangle)$ in mode 2 on a balanced beam splitter $(\hat{B}_{12})$. Since we aim to calculate the variance (signal) and the variance of variance (noise) of the photon number difference detected by camera pixels at position $\vec{x}=(x,y)$, operators $\hat{U}_{1,2}(\vec{x})$ facilitate the basis transformation from the probe and LO beam basis to the pixel basis. The eigenfunction of the $i^{th}$ beam mode, $U_{i}(\vec{x})$, connects the annihilation operators in the beam and pixel basis as $\hat{a}^{\dagger}_{i} = \sum_{\vec{x}}^{} U_{i}(\vec{x})\:\hat{a}^{\dagger}_{\vec{x}}$ \cite{matekole2020quantum,xiao2017hole}. Defining $\hat{N}_{1,2}(\vec{x})$ as the number operator for mode 1 or 2, correspondingly at the output ports, we can write moments of the photon-number difference operator $\hat{\mathscr{N}}(\vec{x})= \hat{N}_1(\vec{x})-\hat{N}_2(\vec{x})$ as:
\begin{equation*}
\expval{\hat{\mathscr{N}}(\vec{x})} = 0,
\label{Eq:mean}
\end{equation*}

\begin{equation}
\begin{split}
&\expval{\hat{\mathscr{N}}^2(\vec{x})} = Tr \Biggl [ \Bigl(\hat{N}_1(\vec{x}) -\hat{N}_2(\vec{x})\Bigr )^{2} \hat{U}_{2}(\vec{x})\hat{U}_{1}(\vec{x})\hat{B}_{12}\\
&\hat{T}_{1}(\vec{x}) \hat{D}_{2}(\alpha)|0 \rangle \langle 0|\hat{\rho}_{1}\hat{D}^\dagger_{2}(\alpha)\hat{T}^\dagger_{1}(\vec{x})\hat{B}^\dagger_{12}\hat{U}^\dagger_{1}(\vec{x})\hat{U}^\dagger_{2}(\vec{x}) \Biggr ].
\label{var1}
\end{split}
\end{equation}
Using intensity of the LO for normalization and by neglecting $\mathcal{O}\left(|\alpha|^{-2}\right)$ terms, it can be shown (Appendix \hyperref[appendix:normvar]{A}) that normalized variance is:
\begin{equation}
  V(\vec{x}) = 1+ 2\expval{\hat{n}_{th}}\left |\tilde{U}_{1}(\vec{x}) \right |^{2}
  \label{eq:variance}
\end{equation}

with $\tilde{U}_{1}(\vec{x}) = U_{1}(\vec{x})\cdot T_{1}(\vec{x})$. This is the normalized variance measured by any individual pixel of the camera and represents the smallest unit of the entire noise statistics. Variance of this variance can be directly calculated with the fact that fourth moment of the Gaussian probability distribution is three times the square of the second moment. Measurement sensitivity can be further improved by effectively increasing the detection area by combining the readout of several nearby pixels, in the process we call "binning". Details of the calculation of binned variance are in Appendix \hyperref[appendix:binning]{B}.

\subsection{Theoretical Signal-to-Noise Ratio (SNR)} 
In differential imaging scheme, object is illuminated with a probe field and its transmitted intensity is compared with that of a reference beam to infer spatial profile of the object. Denoting the mean value of the probe photons as $\expval{\hat{n}}$, theoretical SNR for differential imaging can be defined as
\begin{equation}
SNR = \frac{S_{1}-S_{0}}{\sqrt{\Delta S_{1}^{2}+\Delta S_{0}^{2}}}
\label{SNReq}
\end{equation}
with signal, $S_{0}$ and $S_{1}$ being the detected intensities of the probe and reference beams, respectively. Corresponding noises are quantified by variance terms, $\Delta S_{1}^{2}$ and $\Delta S_{0}^{2}$ with definition, $\Delta S_{i}^{2}=\expval{S_{i}^{2}}-\expval{S_{i}}^{2}$.
To compare our method to a conventional method, we cannot just replace the thermal probe with a laser probe, since this would not produce any image (as the shot-noise variance of a coherent states cannot be distinguished from the shot-noise variance of the vacuum state). The closest alternative is to modify the setup by removing the beam splitter, so that the image is obtained by comparing intensities rather than field amplitudes (in homodyne detection). This is equivalent to the classical differential imaging (CDI) and is often used as a benchmark \cite{brambilla2008high, knyazev2019overcoming, brida2010experimental}. For CDI with coherent state probe of $\expval{\hat{n}_{coh}}$ average photons, Eq.~(\ref{SNReq}) gives
\begin{equation}
SNR_{CDI} = \frac{(1-t)\expval{\hat{n}_{coh}}}{\sqrt{\expval{\hat{n}_{coh}}(1+t)+2(\Delta N_{d}^{2})}}
\end{equation}

where $t$ is the object transmittance and $\Delta N_{d}^{2}$ is the variance of the dark counts. We expect lower SNR if the input state has more noise than a coherent state, e.g., a thermal state \cite{woodworth2020transmission}.
\begin{figure}[ht]
    \centering
    \includegraphics[width=\columnwidth]{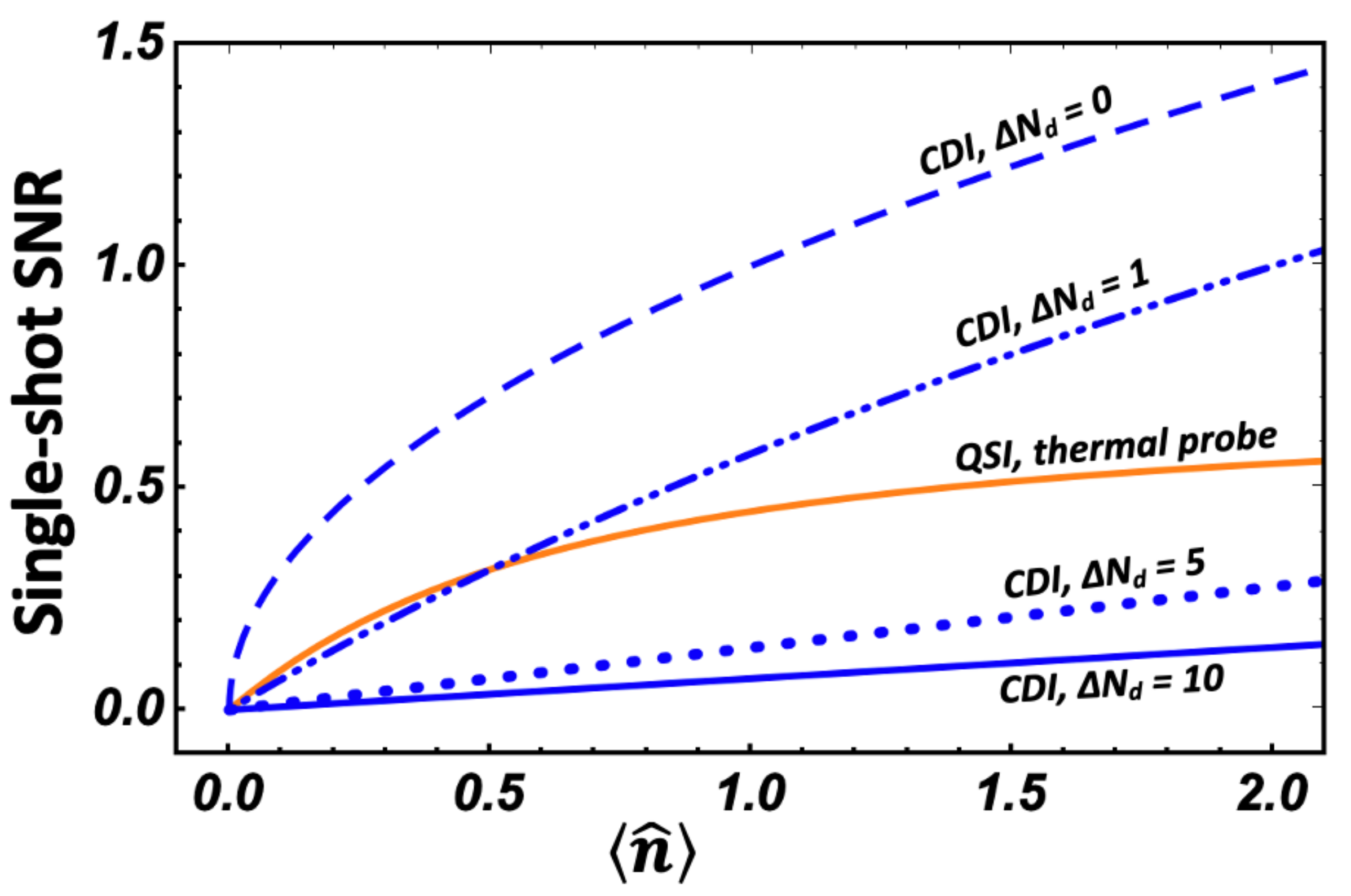}
    \caption {Plot of theoretical single-shot SNR as a function of average probe photons. QSI with thermal probe (orange) is compared with coherent state differential imaging for dark count standard deviation ($\Delta N_{d}$) ranging from 0 to 10 per pixel. For the CCD camera used in our experiments $\Delta N_{d}$ = 10 per pixel. Object is assumed to be completely opaque i.e. $t = 0$.}
    \label{results}
\end{figure}
In the context of QSI, we adapted Eq.~(\ref{SNReq}) by defining signal $S_{0}$ and $S_{1}$ as the variance of the photon number difference in the presence and absence of the object in the setup, respectively. Corresponding `noise' terms, $\Delta S_{1}^{2}$ and $\Delta S_{0}^{2}$, are given by the variance of variance terms. For a thermal probe of $\expval{\hat{n}_{th}}$ average photons, SNR can be calculated as:
 \begin{equation}
SNR_{QSI} = \frac{2 (1-t) \expval{\hat{n}_{th}}}{\sqrt{4+8\expval{\hat{n}_{th}}^2(1+t^{2})+8\expval{\hat{n}_{th}}(1+t)}}
\label{Eq:SNR_Th}
\end{equation}

Fig.~\ref{results} shows comparison in terms of theoretical SNR, between QSI (with thermal state probe) and CDI (with coherent state probe). When dark noise is taken into account, QSI method offers higher SNR than the classical method.

\section{Experimental Method \& Results}
\label{sec:III}
We experimentally demonstrate QSI with weak thermal state, produced using two different sources. Fig.~\ref{Fig:schematic}(c) shows an image of a semi-transparent insect wing, obtained using a pseudo-thermal light, generated by passing a coherent laser field through a rotating diffuser. Corresponding normalized variance map is constructed using 200 image clusters (3 images/cluster, the exposure for each image is $1.7~\mu$s) and with average photon number per pixel per frame $\expval{n}_{pxl} \approx 0.1$, which is well below the dark noise level of $10$ photons per pixel. Thus, this image required total of $60$ photons per pixel. The finer details of the wing structure (width $\approx 17 \mu$m) can be clearly resolved without further processing, as shown in Fig.~\ref{Fig:schematic}(e). However, we can boost the image contrast by increasing the effective number of photons per detected area following the binning process described in Appendix \hyperref[appendix:binning]{B}. Same image, post-processed with the smallest binning radius $R=1$ is shown in Fig.~\ref{Fig:schematic}(f), demonstrates significant improvement in quality, without much deterioration in the spatial resolution. Considering the simplicity of producing such pseudo-thermal light for wide range of optical frequency, this imaging technique can be easily adopted for a broad range of low-light imaging applications where resources like squeezed state are unavailable.

To quantitatively verify theoretical predictions, we switch to weak thermal state generated via the unseeded FWM process. In such a system, a strong pump field interacting with a nearly-resonant atoms in a double-$\Lambda$ configuration results in a strong FWM gain $G$ for a pair of correlated probe/conjugate fields. In case of a seeded probe field, the two-mode intensity squeezed state is produced at the output. In case of vacuum inputs, the output fields are entangled, but each probe and conjugate field individually displays thermal statistics with average photon number $\expval{\hat{n}} = \mathrm{sinh}^2(r)$ and normalized photon count variation $\Delta n^2/\expval{\hat{n}} = 1+\expval{\hat{n}}$. It is convenient to use this source for precise comparison of theory and experiment, since it provides a reliable and repeatable way to control the average photon number in a specific spatial mode through the FWM gain by adjusting the pump beam strength or laser detuning. Moreover, the spatial mode of the thermal state can be directly explored by seeding the input probe channel, and thus we can ensure good spatial overlap between the LO and the thermal field by optimizing the visibility of the interference fringes between the seeded amplified probe field and the LO. Mode matching between thermal and LO modes is quantified by normalized overlap $\mathscr{O}(\vec{x})=\frac{\int_{A}\mathscr{u}_{LO} \mathscr{u}_{Th}^{*} dA}{\sqrt{\int_{A}\mathscr{u}_{LO} \mathscr{u}_{LO}^{*} dA}}$. Normalized variance detected by any pixel at location $\vec{x}$ is affected by the overlap value at that location as $V(\vec{x}) = 1 + 2 \expval{\hat{n}_{th}} t(\vec{x})\cdot \mathscr{O}(\vec{x})$. In the current work, we had overlap close to unity, thus maximizing SNR and ensuring good agreement between theory and experiment. However, poor mode matching will reduce the overlap and deteriorate the SNR.

\begin{figure}[t]
\centering
	\includegraphics[width=\columnwidth]{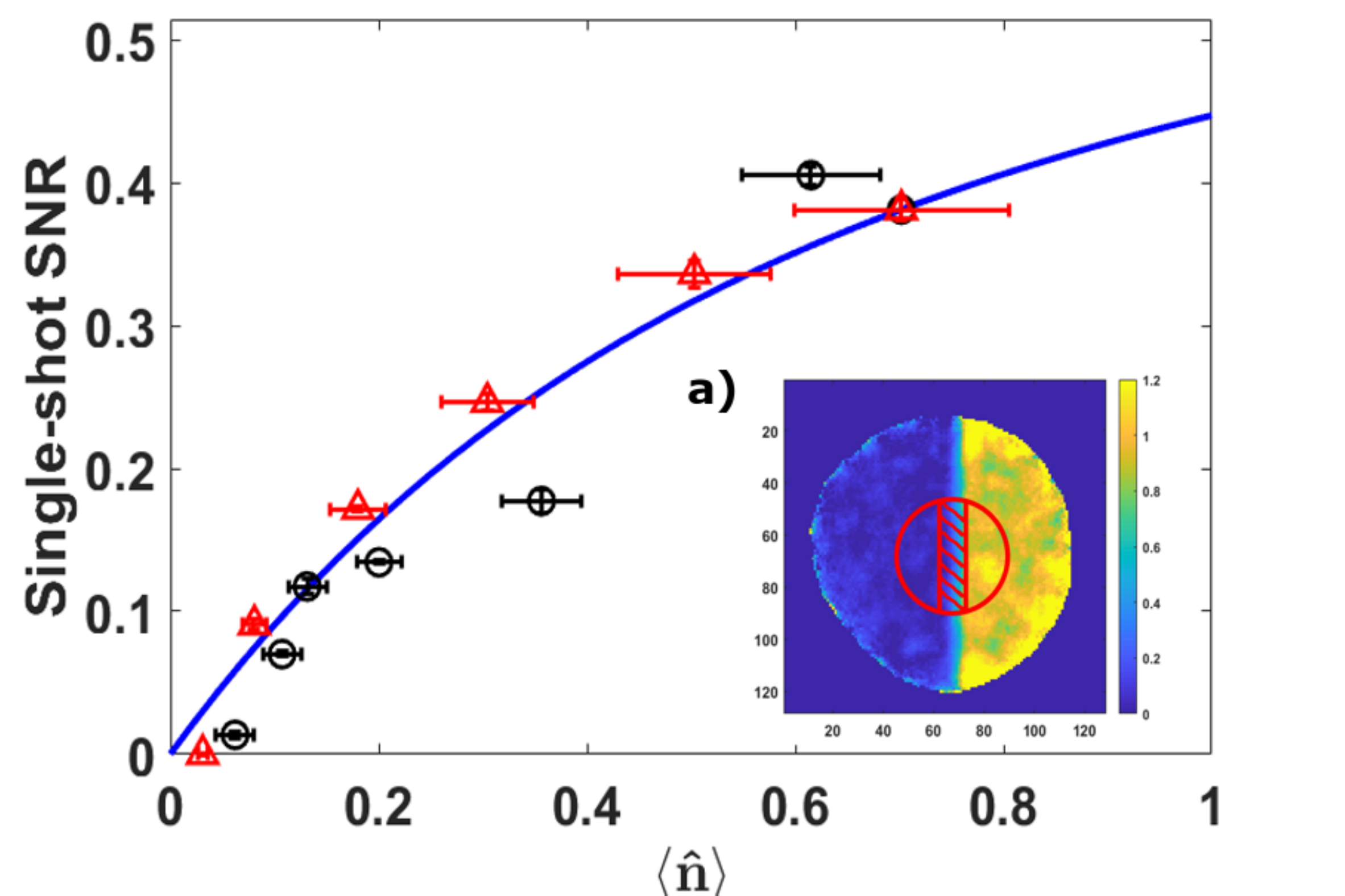}
		\caption{SNR for opaque object imaging in QSI scheme as a function of average thermal photon number per detection area $\expval{\hat{n}}$. The detected photon number was controlled by either changing the total FWM gain (black circles) for detection area size of $a = 113$ pixels, or by changing the detection area size (red triangles) for the highest FWM gain value. Solid blue curve is the theoretical SNR of Eq.~(\ref{correction_equation}). Experimentally measured SNR values are divided by $\sqrt{600}$ for it to be compared with the theoretical values. Inset (a): Transmission map of the opaque object blocking left half of the thermal probe, calculated with the binning radius $6$ ($a$ =113 pixels). A central circular region is selected based on  good overlap between the probe and the LO. Central shaded region is excluded from the analysis.}
	\label{Fig:SNR_Exp}
\end{figure}

We can then recreate the transmission map of the object using the measured quantum noise map as:
\begin{equation}
t = \frac{V_{probe}-1}{V_{ref}-1}, \label{t_V_def}
\end{equation}
where $V_{ref}$ is the normalized quadrature variance of the unobstructed thermal beam and $V_{probe}$ is the normalized quadrature variance measured with object inserted in its path. The procedure of variance map calculation is outlined in Appendix \hyperref[appendix:FWM]{C}. If a portion of the beam is blocked by a completely opaque object, we expect to have on average $t\approx 0$ for the blocked region and $t \approx 1$ for the unblocked region. Inset of Fig.~\ref{Fig:SNR_Exp} is an example of the transmission map of a half-blocked thermal beam.
Since we use $M=600$ image clusters to calculate SNR, it is divided by a factor of $\sqrt{M}$ to compare it to the single-shot theoretical case and ensure fair comparison with .
We compare these experimental SNR measurements with theory by setting $t=0$ in Eq.~(\ref{Eq:SNR_Th}) for a completely opaque object:
\begin{align}
SNR_{QSI}=\frac{2\expval{\hat{n}}}{\sqrt{2+2[1+2\expval{\hat{n}}]^2}}.
\label{correction_equation}
\end{align}
In such case, SNR is determined only by the average number of the detected photons $\expval{\hat{n}}$, and we employ two different strategies to verify that experimentally. One approach is to vary the effective detection area by binning multiple pixels together. The number of photons in detection area can be expressed as $\expval{\hat{n}} = \left(\frac{a}{A}\right )\expval{\hat{n}_{tot}}$, where $a$ is the detection area, $A$ is the whole beam size and $\expval{\hat{n}_{tot}}$ is the total average incoming photon counts. While $a$ is determined by binning radius, $\expval{\hat{n}_{tot}}$ is controlled by FWM gain. For these measurements, we carefully select the `dark' and `bright' areas in a partially blocked thermal beam, shown in the inset of Fig.~\ref{Fig:SNR_Exp}, and verify that for a detection area with less than 10 pixels radius, the recorded optical field is strictly single-mode, and $\expval{\hat{n}}_{pxl}$ is calibrated independently as described in Appendix \hyperref[appendix:calibration]{D}. In second approach, we fixed $a$ and varied $\expval{\hat{n}_{tot}}$ by changing FWM gain with different pump powers. Fig.~\ref{Fig:SNR_Exp} shows the experimental SNR as a function $ \expval{\hat{n}}$ for both methods. Black circles are recorded using fixed detection area and varying FWM gain. Red triangles are recorded by changing detection area while FWM gain is held fixed at its maximum value. Remarkably, both approaches yield experimental data matching very well with each other, proving their equivalence. They are in reasonably good agreement with the theoretical SNR values of Eq.~(\ref{Eq:SNR_Th}). 

In the context of QSI, contrast (C) of the images can be calculated as:
\begin{align}
C=\frac{(V_{bright}-1)-(V_{dark}-1)}{V_{(bright}-1)+(V_{dark}-1)},
\label{contrast}
\end{align}
with $V_{bright}$ and $V_{dark}$ being spatial average of normalized variances recorded by pixels in bright and dark regions, respectively. We measured $C = 0.88 \pm 0.02$ for the image inset in Fig.~\ref{Fig:SNR_Exp}.
% QSI should work with any state, possessing intrinsic noise different from shot noise. Intensity noise of a genuine thermal state generated by using FWM originates from the two-mode quantum noise of twin beams followed by the erasure of the information about one of the modes. The intensity noise produced by a laser field after the RGG, on the other hand, is a result of temporal averaging of many coherent states with different average photon numbers, and thus, is truly classical. Thus, it is not obvious that such a pseudo-thermal light performs as well in QSI method, since its photon statistics approximates the genuine thermal state only in a limited temporal and spatial range. This is why we present both sources, pseudo thermal source in Fig. 1 and genuine thermal source in Fig. S1; to show that both can be used. Then, since they produce the same results, one can use the pseudo thermal state, which is much easier to generate in a wider spectral range.

\section{Conclusion}
\label{sec:IV}
To summarize, we theoretically developed an approach to a low-exposure imaging using classical or quantum state different from a coherent one, and experimentally demonstrated its realization using thermal and pseudo-thermal light. We demonstrated this ability by imaging a biological sample by detecting as low as 0.03 photons/pixel/exposure on average with just 27,000 photons making up the entire image (see Fig.B1 in Appendix \hyperref[appendix:binning]{B}). We also showed that in  the low photon number regime QSI with thermal light outperforms the classical differential imaging method when dark counts are taken into account.

The ability of image reconstruction using very low photon flux is desirable for numerous scientific, commercial, and defense imaging applications. The proposed method offers several attractive features. First and foremost, the wide availability of thermal light sources broadens the scope of applications, potentially extending it in the visible and UV frequencies. Second, since thermal light does not have a set phase, our method does not require the LO phase stabilization, providing substantial increase in reliability. Finally,  since a portion of a thermal field still displays the thermal statistics with lower photon number, the spatial resolution can be optimized depending on the required SNR given by Eq.~(\ref{Eq:SNR_Th}).

\section*{Acknowledgement}
P.B. would like to dedicate this paper to the memory of his advisor, Jonathan P. Dowling. The authors would like to thank Elisha Matekole for useful discussions. This work was supported by the U.S. Air Force Office of Scientific Research through the Grant No. AFOSR FA9550-19-1-0066.

\subsection*{Appendix A: Calculation of Normalized Variance}
\label{appendix:normvar}
%%%%%%%%%% Prefix a "A" to all equations, figures, tables and reset the counter %%%%%%%%%%
\setcounter{equation}{0}
\setcounter{figure}{0}
\setcounter{table}{0}
\makeatletter
\renewcommand{\theequation}{A\arabic{equation}}
\renewcommand{\thefigure}{A\arabic{figure}}
\renewcommand{\bibnumfmt}[1]{[#1]}
\renewcommand{\citenumfont}[1]{#1}
%%%%%%%%%% Prefix a "A" to all equations, figures, tables and reset the counter %%%%%%%%%%
In Eq.~(1), order of intra-mode and intermode interferences can be exchanged (i.e. $\hat{U}_2\hat{U}_1\hat{B}_{12} = \hat{B}_{12}\hat{U}_2\hat{U}_1$). Using this, along with cyclic property of trace, Eq.~(1) can be written as:
\begin{equation}
\begin{split}
\mathscr{V}(\vec{x})= & Tr \Biggl [\hat{B}^\dagger_{12} \left(\hat{N}_1(\vec{x}) -\hat{N}_2(\vec{x})\right )^{2}\hat{B}_{12}\hat{U}_{2}(\vec{x})\hat{\tilde{U}}_{1}(\vec{x})\\
&\hat{D}_{2}(\alpha)\left|0 \rangle \langle 0\right|\hat{\rho}_{1}\hat{D}^\dagger_{2}(\alpha)\hat{\tilde{U}}^\dagger_{1}(\vec{x})\hat{U}^\dagger_{2}(\vec{x})\Biggr ]
\label{var2}
\end{split}
\end{equation}
where $\hat{\tilde{U}}_{1}(\vec{x}) = \hat{U}_{1}(\vec{x})\cdot \hat{T}_{1}(\vec{x})$. For any unitary operators $\hat{P}$ and $\hat{Q}$,
\begin{align}
&\hat{P}^{\dagger} \hat{Q}\hat{P} \rightarrow  P^{-1}\hat{Q}\nonumber \\
&\hat{P}^{\dagger} \hat{Q}^\dagger \hat{P} \rightarrow  (P^{-1})^{*}\hat{Q}^\dagger,
\label{unitaryM}
\end{align}
with $P^{-1}$ being the inverse of the matrix representation of $\hat{P}$. This leads to $\hat{B}^\dagger_{12} \left(\hat{N}_1(\vec{x}) -\hat{N}_2(\vec{x})\right )^{2}\hat{B}_{12} = \hat{a}_{1}^\dagger \hat{a}_{1} + \hat{a}_{2}^\dagger \hat{a}_{2}+ 2\hat{a}_{1}^\dagger \hat{a}_{1}\hat{a}_{2}^\dagger \hat{a}_{2}-\hat{a}_{1}^{2} \hat{a}_{2}^{\dagger^{2}}-\hat{a}_{2}^{2} \hat{a}_{1}^{\dagger^{2}}$ and 
Eq.~(\ref{var2}) yields:
\begin{equation}
\begin{split}
\mathscr{V}(\vec{x}) = &\left |U_{2}(\vec{x}) \right |^{2} \left |\alpha \right |^{2}+\expval{\hat{n}_{th}} \left |\tilde{U}_{1}(\vec{x}) \right |^{2}+\\ & + 2\expval{\hat{n}_{th}}\left |\alpha \right |^{2}\left |U_{2}(\vec{x}) \right |^{2}\left|\tilde{U}_{1}(\vec{x}) \right |^{2}.
\label{var3}
\end{split}
\end{equation}
Using intensity of the local oscillator for normalization and by neglecting $\mathcal{O}\left(|\alpha|^{-2}\right)$ terms, normalized variance is:
%\begingroup
%\setlength{\belowdisplayskip}{0pt}
\begin{equation}
   V(\vec{x}) = 1+ 2\expval{\hat{n}_{th}}\left |\tilde{U}_{1}(\vec{x}) \right |^{2}.
\end{equation}

\subsection*{Appendix B: Binning}
\label{appendix:binning}
\setcounter{equation}{0}
\setcounter{figure}{0}
\setcounter{table}{0}
\makeatletter
\renewcommand{\theequation}{B\arabic{equation}}
\renewcommand{\thefigure}{B\arabic{figure}}
\renewcommand{\bibnumfmt}[1]{[#1]}
\renewcommand{\citenumfont}[1]{#1}

In a CCD camera, each pixel acts as an independent detector, collecting only the light falling on its surface. Since the mode size of thermal field is much larger than the pixel size (13 $\mu m \times 13 \mu m$), the average number of photons per pixel $\langle n \rangle_{pxl}$ is proportionally small, and the variance value is close to one, making it hard to distinguish from the coherent vacuum. To improve the sensitivity of our measurements, we group pixels together to effectively increase their cumulative detection area. In our binning protocol, individual photon count of each pixel at $\vec{x}$ is replaced by the sum of photon counts of all the neighbouring pixels within a binning radius, $R$. Binning improves the SNR but at the cost of reduced spatial resolution.

Since we include all the pixels within the binning radius i.e. $|\vec{x}-\vec{x}^{'}|\leq R$, binned variance can be written as:
\begin{equation}
\begin{split}
\mathscr{V}_{R}(\vec{x}) = & Tr\Biggl[ \hat{\tilde{U}}^\dagger_{1}(\vec{x})\hat{D}^\dagger_{2}(\alpha)\hat{U}^\dagger_{2}(\vec{x})\hat{B}^\dagger_{12}\sum_{\vec{x}^{'}}^{}\left(\hat{N}_1(\vec{x}^{'}) -\hat{N}_2(\vec{x}^{'})\right)^{2}\\
&\hat{B}_{12}\hat{U}_{2}(\vec{x}^{'})\hat{D}_{2}(\alpha)\hat{\tilde{U}}_{1}(\vec{x}^{'})|0 \rangle \langle 0|\hat{\rho}_{1} \Biggr ].
\end{split}
\end{equation}
 Central sum of the previous equation can be written as a product of two terms,
\begin{equation}
\begin{split}
&\left\langle \left(\sum_{\vec{x}^{'}}\hat{N}_1(\vec{x}^{'}) -\hat{N}_2(\vec{x}^{'})\right)^2 \right \rangle =\sum_{\vec{x}^{'}}\left \langle\left(\hat{N}_1(\vec{x}^{'}) -\hat{N}_2(\vec{x}^{'})\right)^2\right \rangle\\
&+\sum_{\vec{x}^{'}}\sum_{\vec{x}^{''}\neq \vec{x}^{'}}\left\langle \left(\hat{N}_1(\vec{x}^{'}) -\hat{N}_2(\vec{x}^{'})\right) \left(\hat{N}_2(\vec{x}^{''}) -\hat{N}_2(\vec{x}^{''})\right) \right\rangle,
\end{split}
\end{equation}

in which the first term is already evaluated in Eq.~(\ref{var3}) and the second term can be evaluated using Eq.~(\ref{unitaryM}), to get the binned variance:
\begin{equation}
\begin{split}
&\mathscr{V}_{R}(\vec{x})= \left |\alpha \right |^{2}\sum_{\vec{x}^{'}}^{}\left |U_{2}(\vec{x}^{'}) \right |^{2} \left (1 + 2\expval{\hat{n}_{th}}\left|\tilde{U}_{1}(\vec{x}^{'}) \right |^{2}\right)\\
&+2\expval{\hat{n}_{th}}|\alpha|^{2}\left(\left |\sum_{\vec{x}^{'}}^{} U^{*}_{2}(\vec{x}^{'})\tilde{U}_{1}(\vec{x}^{'})\right|^{2}- \sum_{\vec{x}^{'}}^{}\left | U^{*}_{2}(\vec{x}^{'})\tilde{U}_{1}(\vec{x}^{'})\right|^{2}\right ).
\label{var4}
\end{split}
\end{equation}
$T_{1}$ matrix is written as a diagonal matrix with entries being 0's (1's) representing the presence (absence) of the opaque part of the object. Mode matching between probe and LO allows us to write $\tilde{U}_{1}(\vec{x}) = T_{1}(\vec{x})\cdot U_{2}(\vec{x})$. With this, expression for binned normalized variance simplifies to
\begin{equation}
V_{R}(\vec{x})= 1 + 2\expval{\hat{n}_{th}}\frac{\left(\sum_{\vec{x}^{'}}T_{1}(\vec{x}^{'})\left|U_{2}(\vec{x}^{'})\right|^{2}\right)^{2}}{\sum_{\vec{x}^{'}}^{}\left |U_{2}(\vec{x}^{'}) \right |^{2}}.
\label{var5}
\end{equation}

\begin{figure}[h]
\centering
	\includegraphics[width=\columnwidth]{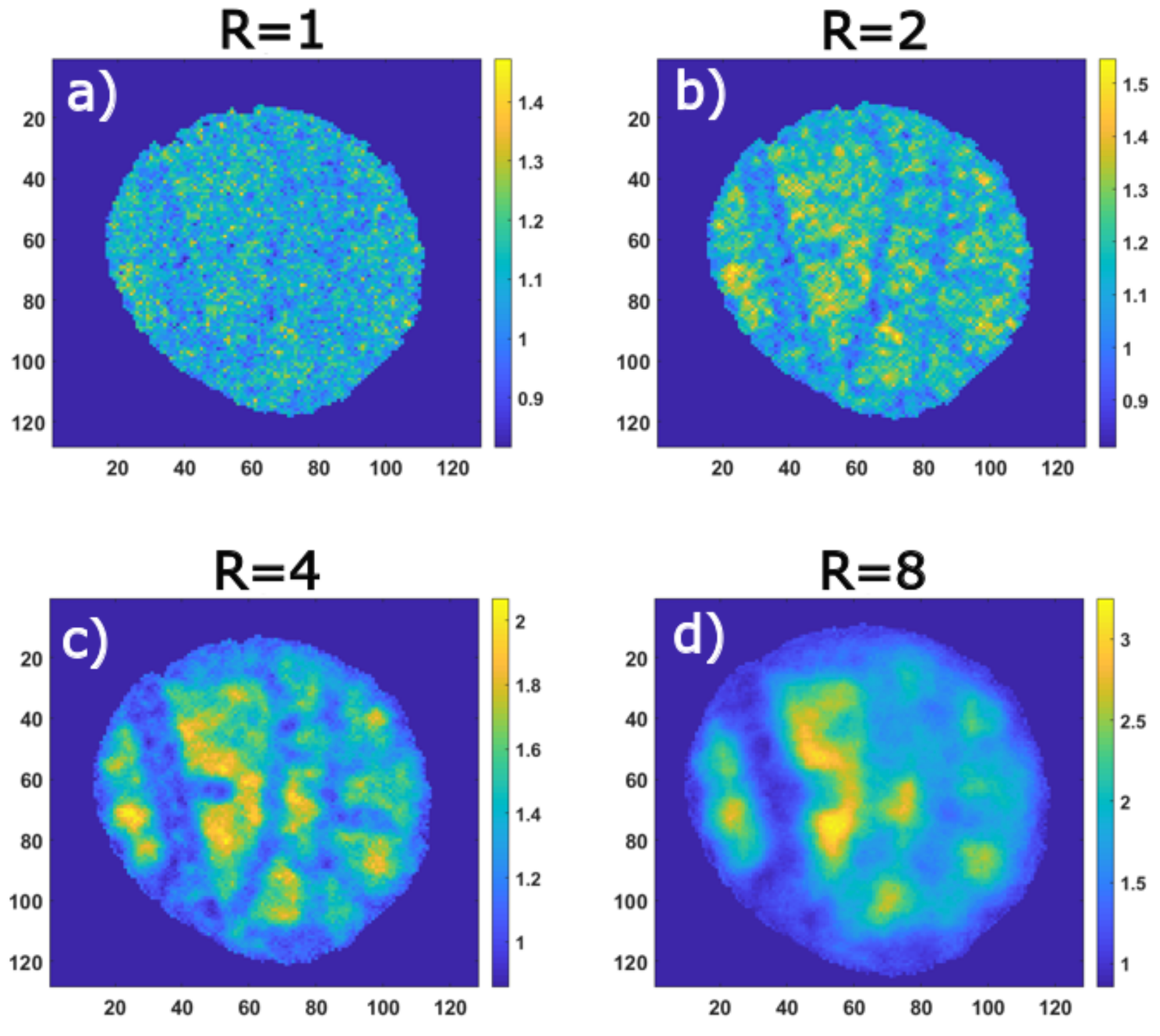}
	\caption{Variance map (a) of the insect wing is generated with $R =1$ and 0.03 photons per pixel per 1.7~$\mu$s exposure. Images (b-d) are generated with different values of the binning radii of $R = $2, 4, and 8.}
	\label{compare}
\end{figure}

Fig.~\ref{compare} shows variance maps of an insect wing at different levels of binning. To construct this variance map, we recorded 0.006 photons/pixel/frame (total of $\approx$ 27,000 thermal photons with 600 frames, generated by the FWM method). See Section \ref{appendix:calibration} for photon number calibration. Since the process of binning combines photon counts of all the neighbouring pixels in binning radius R, the `effective' photon/pixel/frame count to generate e.g. Fig. S1 (a) is $0.006\times 5$. Since the variance of the coherent field in the blocked region and that of an unobstructed weak thermal beam are not very different, smaller bin sizes yield low contrast. Higher binning radii yield better contrast between the two regions, as shown in the last row of Fig.~\ref{compare}, although their boundary is smoothed by the binning process thus degrades resolution.

\subsection*{Appendix C: Detailed Experimental Procedure}
\label{appendix:FWM}
\setcounter{equation}{0}
\setcounter{figure}{0}
\setcounter{table}{0}
\makeatletter
\renewcommand{\theequation}{C\arabic{equation}}
\renewcommand{\thefigure}{C\arabic{figure}}
\renewcommand{\bibnumfmt}[1]{[#1]}
\renewcommand{\citenumfont}[1]{#1}
All the imaging data are recorded using the Princeton Pixis 1024 CCD camera, which has low dark noises counts (standard deviation $\approx$ 10 dark counts per pixel) and high quantum efficiency ($95$\%). To work around the low speed of the camera, we record a quick sequence of images that form a cluster, with 1.7~$\mu$s exposure time for a 544 $\mu$s duty cycle. Each cluster consists of 6 images, although only the second through fourth images are suitable for the analysis due to leakage contamination. Further steps are shown in Fig.~1(b): recorded images are spatially matched and subtracted pairwise across the beam region (similar to the traditional homodyne detection). We then calculate normalized temporal variance within one cluster (6 images total, 3 matched pairs) as:
\begin{equation}
V(\vec{x}) = \frac{\big\langle  \big( N_1 (\vec{x}) - N_2 (\vec{x}) \big)^2 \big\rangle}{\big\langle N_1 (\vec{x}) + N_2 (\vec{x})\big\rangle}. \label{variance_exp}
\end{equation}
Here we assume that the two recorded beams have identical spatial distributions and are perfectly matched. $N_1 (\vec{x})$ and $N_2 (\vec{x})$ are photon counts at the point $\vec{x}$ within corresponding beams. $V(\vec{x})$ is further averaged over all the kinetic clusters in a given set to obtain the final variance map.

\begin{figure}[h]
\centering
	\includegraphics[width=\columnwidth]{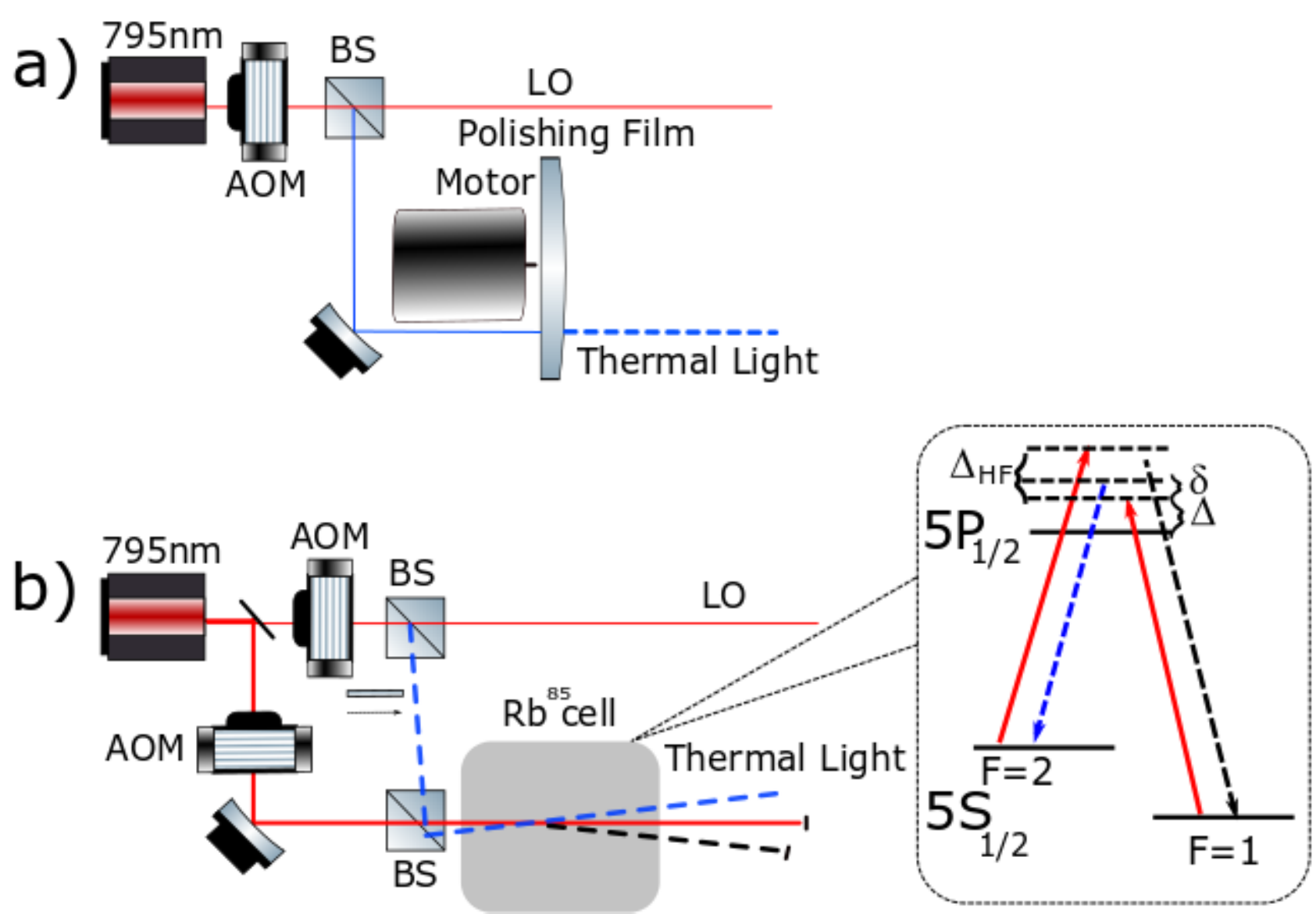}
	\caption{Experimental Setup for weak thermal state generation. (a) Pseudo-thermal state generated by scattering a coherent beam with rotating polishing film with grain size $\approx 0.1\mu m$. b) Thermal state source using FWM process: vacuum input is used as seed to generate weak thermal state. Inset: atomic level diagram of FWM process. $\Delta_{HF}$ is the hyperfine splitting between $5S_{1/2} F=2$ and $5S_{1/2} F=1$ level. $\Delta$ is the one-photon detuning and $\delta$ is the two-photon detuning. The red line stands for the pump laser. The blue and black lines represent probe and conjugate beams, respectively~\cite{prajapati2019polarization,prajapati2019optical}. }
	\label{Fig:setup}
\end{figure}

We can construct the transmission map of the object from the variance map using Eq.~(6). Experimentally, the SNR is calculated using the transmission map by selecting two regions of interest (ROI), each within the blocked and unblocked halves (see inset of Fig.~4. We selected only small central area with proper spatial overlap between the probe beam and LO. With this, experimental SNR is given by,
\begin{equation}
SNR_{single} = \frac{S_{unblocked}-S_{blocked}}{\sqrt{\Delta S_{unblocked}^2+\Delta S_{blocked}^2}\sqrt{M}}
\label{snr_eqn}
\end{equation}
with $ S_{unblocked} $ and $S_{blocked}$ being the means and $\Delta S^2_{unblocked}$ and $\Delta S^2_{blocked}$ being the spatial variances of the two ROI in the blocked and unblocked halves of the transmission map. $M$ is the total number of datasets.
Fig.~\ref{Fig:setup} shows two different thermal sources involved in this work. Fig.~\ref{Fig:setup}(a) shows the rotating diffuser setup. A coherent input field ($\lambda\approx 795$ nm) is sent through a rotating diamond polishing film with grain-size 0.1 $\mu m$. Both, the probe beam and the LO are pulsed using acousto-optical modulator (AOM) synchronized with the camera's image-taking sequence to avoid contamination due to prolonged exposure and to reduce 1/f camera noise.

Schematic of the FWM thermal source is shown in Fig.~\ref{Fig:setup}(b). Pump field ($\lambda = 794.7930$~nm) is collimated (beam diameter  0.55 ~mm) and directed to a $2.5$~cm-long $^{85}$Rb vapor cell, maintained at $104.5^{\circ}$C. A fraction of the same laser output is used to produce both, LO at probe field frequency and the input probe field, when necessary. For that, the split beam is phase-modulated at $3035$~MHz (corresponding to the $^{85}$Rb $5S_{1/2}$ hyperfine splitting) using a fiber electro-optical modulator (fEOM), and the lower modulation sideband is filtered using a Fabry-Perot (FP) etalon. Under these conditions, the FWM gain is sufficiently strong $G\ge7$ for $\approx 80$~mW pump power, with no significant atomic absorption at the probe field frequency. After the vapor cell, the pump field is filtered using the polarization and spatial filtering, and only the output probe field is directed first to the imaged object and then to the detection unit. Both, the pump and the probe fields are pulsed using AOMs synchronized with the camera's image-taking sequence.

\subsection*{Appendix D: Photon Number Calibration }
\label{appendix:calibration}
\setcounter{equation}{0}
\setcounter{figure}{0}
\setcounter{table}{0}
\makeatletter
\renewcommand{\theequation}{D\arabic{equation}}
\renewcommand{\thefigure}{D\arabic{figure}}
\renewcommand{\bibnumfmt}[1]{[#1]}
\renewcommand{\citenumfont}[1]{#1}
\begin{figure}[ht]
\centering
	\includegraphics[width=\columnwidth]{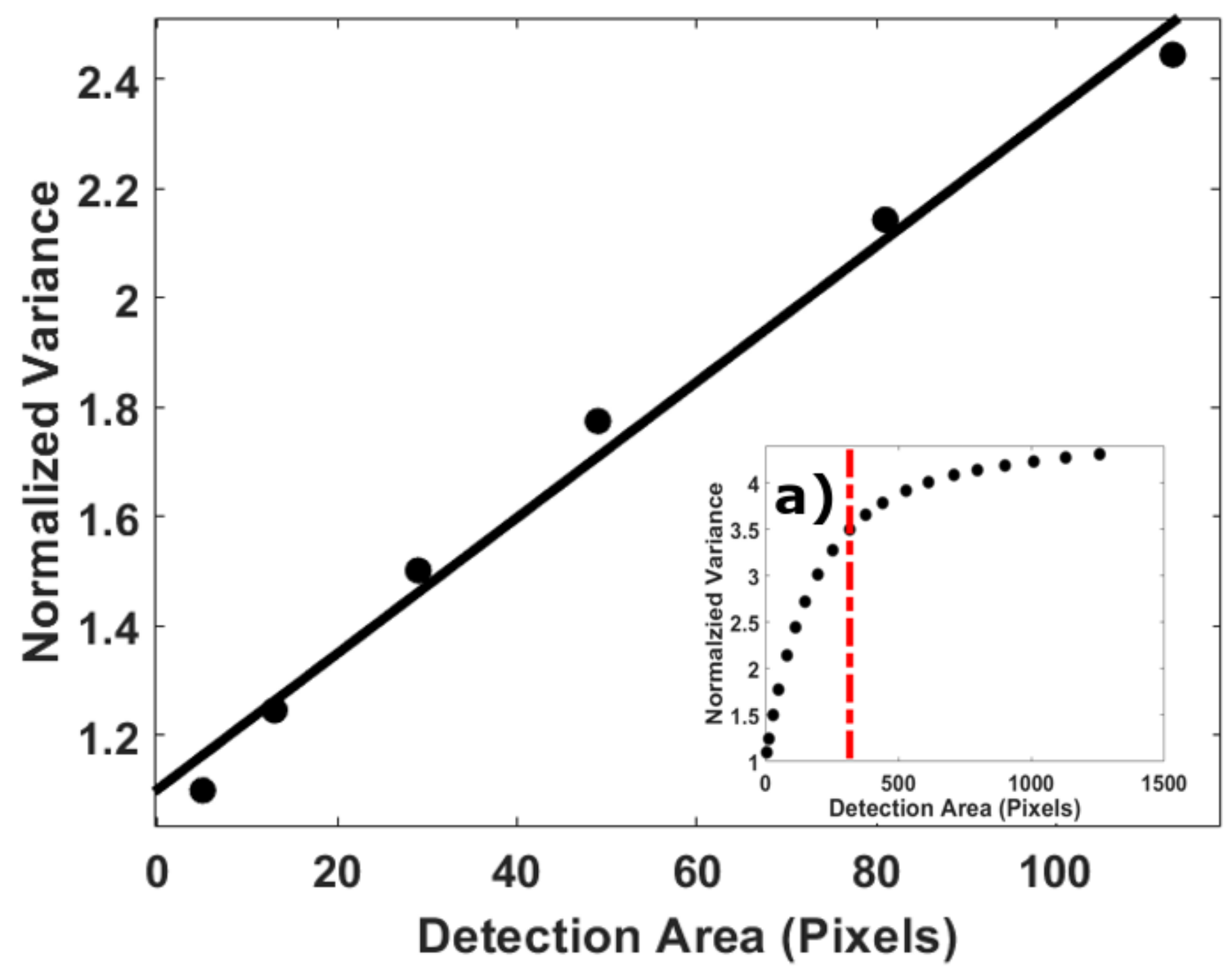}
	\caption{Normalized variance ($V$) as a function of detection area $(a)$. The best linear fit is $V = a \times( 0.01242 \pm 0.0018) + (1.10\pm 0.11)$, matching the predictions of Eq.~(2). Inset: same plot for wider range of detection areas, showing saturation of $V$ for increasing $a$.}
	\label{ModeSize}
\end{figure}
To compare the experimental SNR values with Eq.~(5), we need to accurately estimate the average photon number in the thermal probe field. We need to take into account the bin size and the number of physical pixels integrated during the binning process. If $\langle n\rangle_{pxl}$ is the average number of photons in the unobstructed thermal beam per physical pixel then the number of photons in each binned pixel scales as $\expval{\hat{n}}=a \expval{\hat{n}}_{pxl}= \frac{a}{A}\expval{\hat{n}_{tot}}$. Conveniently, this allows us to vary average photon number by varying the binning radius. Since our measurement procedure involves subtraction of the two outputs before the camera, the measured normalized variance of the photon counts as a function of transmission coefficient is given by Eq.~(2), where it maps to $\abs{\tilde{U}_{1}(\vec{x})}^2$.
Fig.~\ref{ModeSize} shows the measured normalized variance as a function of photon number (black dots) for a set of experimental data obtained with FWM method. As expected, variance depends linearly on the bin area for small binning radii. The slope of this curve allowed us to extract information about $\expval{n}_{pxl}$ = 6.2$\times 10^{-3}$ for pump power $\approx$ 80 mw. However, for large binning ($R>10$), variance starts to deviate from the linear behavior and exhibits signs of saturation (Fig.~\ref{ModeSize} inset). This behavior can be explained by noting that the outputs of the FWM process are expected to contain multiple spatial modes~\cite{kumar2019spatial,holtfrerich2016control}. Assuming $j$ such thermal modes to be equally populated, theoretical normalized variance saturates as $V = 1+\frac{2\:\expval{\hat{n}_{th}}}{j}$ \cite{mandel1995optical}. Thus, for a larger binning radii, measured variance must contain contributions from multiple thermal modes, deviating from the predictions of a single-mode theory.

\setcounter{equation}{0}
\setcounter{figure}{0}
\setcounter{table}{0}
\makeatletter
\renewcommand{\theequation}{\arabic{equation}}
\renewcommand{\thefigure}{\arabic{figure}}
\renewcommand{\bibnumfmt}[1]{[#1]}
\renewcommand{\citenumfont}[1]{#1}

\bibliography{References.bib}
\bibliographystyle{unsrt}
% \bibliography{bibliography.bib}
% \bibliographystyle{osajnl}
%abbrv, unsrt, apha, acm, agsm, vancouver, apalike

\end{document}